Integration of AI-Driven CAD Systems in Designing Water and Power Transportation

Infrastructure for Industrial and Remote Landscape Applications

Sunggyu Park

University of Central Missouri

INDM 5230 Seminar IM

Dr. Suhansa Rodchua

Dec 07, 2025

**Abstract**

The integration of AI into CAD systems transforms how engineers plan and develop infrastructure projects involving water and power transportation across industrial and remote landscapes. This paper discusses how AI-driven CAD systems improve the efficient, effective, and sustainable design of infrastructure by embedding automation, predictive modeling, and real-time data analytics. This study examines how AI-supported toolsets can enhance design workflows, minimize human error, and optimize resource allocation for projects in underdeveloped environments. It also addresses technical and organizational challenges to AI adoption, including data silos, interoperability issues, and workforce adaptation. The findings demonstrate that AI-powered CAD enables faster project delivery, enhanced design precision, and increased resilience to environmental and logistical constraints. AI helps connect CAD, GIS, and IoT technologies to develop self-learning, adaptive design systems that are needed to meet the increasing global demand for sustainable infrastructure.

## Table of Contents

**Introduction**

The demand for infrastructure is increasing as technology evolves, requiring greater complexity and multiple software applications, especially in remote and industrial regions where access to utilities like water and power is essential for industrial operations, construction, or demolition. The need for automated and advanced planning tools in civil design has never been greater, due to the urbanization towards satellite towns and expansion towards remote areas. The development of artificial intelligence (AI) enables software engineers to explore potential areas that could be highly useful by integrating AI with other software, such as computer-aided design (CAD) and geographic information systems (GIS). The integration of AI with CAD can create transformative synergy across many industries, including how infrastructure is conceptualized, analyzed, and constructed. A recent industry dataset also shows measurable improvements from AI-enhanced design tools. A 2025 industry report found that architecture and engineering firms using AI-powered CAD and BIM systems achieved approximately 30–50% efficiency gains through intelligent automation and design optimization (Monograph, 2025).

Civil engineering was designed by hand for many years in the 20th century, and CAD revolutionized the technical drawing industry in the late 20th century. This technical breakthrough in technical drawing expanded the design and drafting industry, reaching far beyond the limitations of traditional technical drawing. The importance of CAD in civil engineering has been well established throughout its history, particularly in the design of roads and pipelines. However, conventional CAD tool methods had several limitations, requiring extensive manual input and rule-based modeling. These limitations could reduce the efficiency and adaptability of CAD tools. Enhancing the CAD process with AI can introduce machine learning algorithms, generative design, and data-driven optimization. These enhancements will open new opportunities for CAD software and provide a strong direction for the next generation of CAD, enabling CAD experts to simulate real-world scenarios, accelerate the design process by reducing design iterations, and enhance the overall efficiency of the design process.

### Purpose of the Study

The paper examines how integrating AI into CAD systems can revolutionize the CAD process and its outcomes, particularly in water and power transportation within construction zones or mining areas in underdeveloped environments. Access to water and power is essential in construction zones and mining areas, but obtaining them can be very challenging, especially in underdeveloped, remote areas. The current issues in infrastructure design, including the evaluation of advantages and challenges of integrating CAD and AI, as well as business and ethical implications, will be discussed. Additionally, insights into predicting future trends will be provided. Ultimately, the discussion of these possibilities will pave the way for further innovations that could reshape CAD in civil infrastructure planning to meet modern demands for efficiency, sustainability, and resilience.

### Background

The new development continues to expand from the edge of the urban area into an undeveloped area. The demand for new development in remote regions to support mining operations, the creation of satellite towns, and the expansion of urban environments is expected to continue increasing (Liu S, 2023, p. 13). As demand for urbanization increases, the demand for efficient water and power transportation infrastructure also rises. Water and power delivery are essential sectors with complex operational requirements that require robust design, resilience, and sustainability. These requirements become even more challenging when conducted under remote and underdeveloped worksites. Conventional approaches to civil design relied heavily on manual, labor-intensive CAD drawing. Due to these limitations, the industry has historically struggled to meet these demands, and these demands are increasing as challenges persist. Due to these limitations and challenges, the traditional approach to civil design often results in inefficiencies, protracted design timelines, and limited flexibility to adapt to environmental and logistical constraints.

It becomes very challenging to implement modern infrastructure in a remote area, as it faces limited resource optimization, the need for minimal environmental impact, and the need to lower lifecycle costs while ensuring operational safety. The integration of geographic information systems (GIS), the Internet of Things (IoT), and building information modeling (BIM) has significantly enhanced decision-making and created numerous opportunities for CAD experts. Yet, constraints persist in adaptive, data-driven project solutions (Liu X, 2025, p. 14). Building Information Modeling has been implementing its workflow with AI, and many mechanical design software, like SolidWorks and Inventor, have been using AI for calculations for many years. Recently, all these software programs have introduced design model suggestions that calculate many different factors designers need to consider (Kutá & Faltejsek, 2025, p. 4).

With the recent advancement in artificial intelligence (AI), particularly in machine learning, predictive analysis, and generative design, these aspects of AI hold significant potential to change the CAD landscape. By integrating AI algorithms with CAD tools, experts can rely on scenario testing, optimization, and simulated feedback to streamline and plan the design process.

The integration of disciplines in construction, mining, and civil design requires multidisciplinary experts to address the unique characteristics of the work site and respond dynamically to adjust project parameters, regulations, or environmental risks.

### Problem statements

Several challenges could arise in industrial, construction, and mining sites, despite the evident potential of AI-driven CAD systems. These challenges include the complexity of requirements, manual limitations, data silos, resource inefficiencies, and technological adoption barriers.

As modern projects are often conducted in remote areas, the complexity of project requirements increases with scale, due to distinct environmental, regulatory, and topographical conditions associated with larger projects. Many rural projects are also site-specific and require multidisciplinary teams with different specializations due to the unique characteristics of the project (Graham, 2023, para. 13). Even with the assistance of AI, some manual input from a mechanical designer is still required. There will always be manual limitations because the manual input always leaves room for human error or oversight in critical aspects. This can be more critical to the process because a small mistake could lead to a larger error. Also, misunderstandings between AI and designers could lead the design in the wrong direction.

Data silo are one of the biggest challenges software integrations face because many software systems are not compatible with each other. If there are no solutions for the data silo, there could be even more manual work needed, even with the use of AI. Data should flow seamlessly between designing, monitoring, and analysis (U.S. CAD, 2025, Para. 3).

Resource inefficiencies are one of the most critical challenges facing power and water transportation. Resource inefficiencies become even more challenging when a project is conducted in a remote area with minimal or aging infrastructure, or no infrastructure access at all. In the United States of America, there is around 14% to 18% loss of treated drinking water every day, due to aged infrastructure (Aurigo, 2025, Para. 5). This problem still exists because it will cost billions of dollars for the government to resolve this problem. This illustrates the extent to which potable water infrastructure is unreliable in remote areas.

The technological adoption barrier can also impact the integration of AI with CAD, due to concerns regarding resistance to interoperability, cost, workforce upskilling, and data security. These challenges underscore the need for integrated, adaptive, and intelligent approaches, particularly in remote environments where traditional resources and methods are insufficient. AI-driven CAD can open numerous opportunities; however, the complexity and limitations in remote, underdeveloped, and logistically challenging areas pose significant challenges. Overcoming these problems requires technological innovation, as well as organizational and systemic changes, with improved data interoperability, adaptable infrastructure, and workforce development. These challenges are crucial to understanding the full potential of an AI-integrated design system in enhancing the efficiency, resilience, and sustainability of water and power transportation infrastructure in remote and underdeveloped areas of construction and mining.

Diagram 1, Factors - Barriers

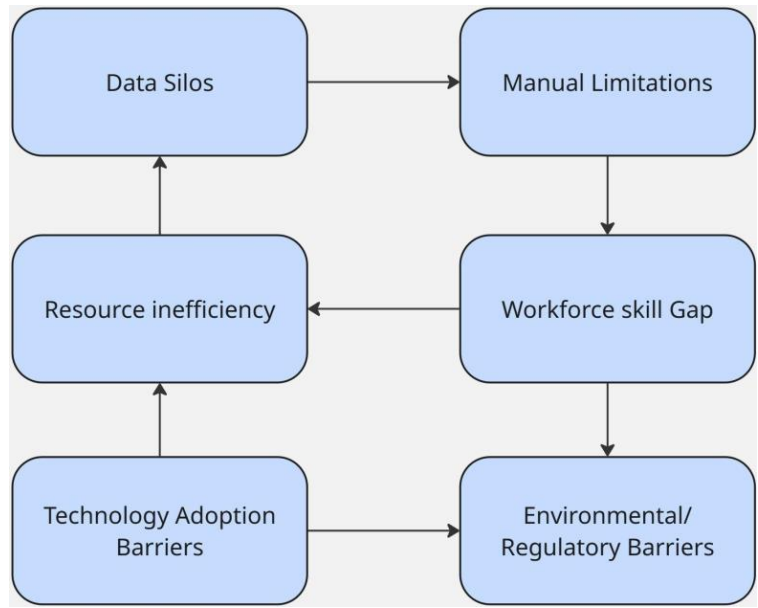

**Methodology**

To analyze the integration of AI-driven CAD systems in infrastructure design, a qualitative literature review will be adopted. Recent peer-reviewed journals, conference proceedings, and professional engineering databases were carefully selected to provide accurate insight. This paper focuses on AI applications in water and power transportation in remote and industrial environments. A literature review is a well-established and effective method in engineering and technology research. It identifies knowledge gaps, allows for new findings, and evaluates emerging trends (Snyder, 2024, pp. 110-113). It draws on sources from academic journals such as Automation in Construction, Applied Sciences, and Fluid Science and Engineering, as well as real-life technical reports from various organizations, including the U.S. government, Aurigo, and PlanRadar. These sources are often from open-access journals, such as those found on Google Scholar, ScienceDirect, and SpringerLink. Source selection is based on the study's focus on AI integration in CAD, its application in water and power transmission, and infrastructure challenges in remote and rural areas. Recent studies in engineering design emphasize the potential value of systemic and integrative reviews for connecting technological development with practical industry applications (Escudero-Mancebo et al., 2023, p. 45).

Diagram 2, Research Process

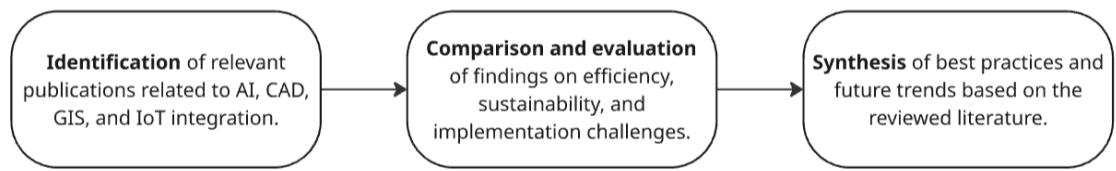

This systematic evaluation approach will assess both the technological potential and practical barriers of AI-assisted CAD systems while maintaining academic insights in real-world engineering applications. The cross-disciplinary analysis will connect the AI research with industrial design, data management, and civil design, especially in fluid infrastructure.

To ensure methodological rigor, the literature search adheres to clear inclusion and exclusion criteria, targeting English-language publications from the last five years that address AI-driven CAD in infrastructure. The keywords include "AI," "Computer-Aided Design," "infrastructure," "water transportation," and "power transportation," among others, and are used in combination. The sources are drawn from multiple databases to ensure comprehensive search results. After eliminating duplicate entries and results not relevant to the context, the final sample is selected based on thematic depth, relevance, and quality.

Thematic analysis synthesizes findings by identifying key trends, challenges, and innovations through categorization into themes related to technological advances, implementation barriers, and future directions. Divergent opinions and cross-disciplinary insights are discussed and tabulated to highlight those areas that represent consensus and those that reflect debate. Methodological transparency is assured by documenting each step, enabling replication and facilitating future systematic reviews in this domain.

**Applications in water transportation**

Efficient and resilient water transportation system design in remote and underdeveloped areas requires high precision planning, logistical accuracy, and long-term sustainability. The ability to address the persistent challenges in terrain adaptability, demand fluctuations, and lack of infrastructure through algorithms and real-time designing capabilities.

Water transportation in mining zones is most likely to require a custom layout to accommodate challenging topographies and potential obstacles in both urban and undeveloped areas. AI generative design can support design professionals by evaluating numerous possibilities and suggesting optimal configurations, considering energy and material consumption, as well as environmental impact (Zhou M, 2022, p. 7). Machine learning algorithms can predict pipeline performance and the risk of infrastructure failure, helping design experts mitigate maintenance challenges and abnormal pressure.

AI-integrated CAD systems can extract real-time data from geographic information technology and Internet of Things sensors to make dynamic adjustments to design parameters, such as flow rate, distance, plumbing requirements, and contingency planning. For example, AI could simulate augmented reality in CAD and suggest design corrections or adaptive operating strategies by testing different demand or irregular flow patterns (Banihashemi et al., 2024, p. 3). Having AI design the flow direction and assign tag numbers would be very convenient for mechanical designers in pipe design.

AI can train predictive models using historical failure data and apply them to improve predictive accuracy, thereby reducing unscheduled downtime and leakage risks. These supervisions and maintenance tasks can be facilitated by AI, which can help identify resource inefficiencies, especially in remote regions. This could revolutionize the process in rural and underdeveloped areas, where it currently requires extensive manual analysis and intervention that logistical constraints can delay.

**Application for power transportation**

Power transportation could be even more impactful than water transportation in many situations. The reliability and efficiency of the power transportation are top priorities because, without power, most operations are impossible. The power outage could not only halt operations but also pose safety risks. With AI-driven CAD, we can transform how power infrastructures, such as transmission lines, substations, and distribution networks, are planned, modeled, and managed.

AI-driven CAD analyzes various terrain and environmental datasets to analyze and optimize route planning. Key datasets, such as topology, vegetation, climatic risks, and proximity to existing structures, can inform optimal power transportation routes. These datasets will be used again to evolve machine learning algorithms to evaluate thousands of power transportation permutations, considering multiple factors such as cost, material constraints, environmental regulations, safety factors, and maintenance access. Ultimately, the optimization will reduce final cost, time, and operational risks (Ajao, 2025, P. 3523).

AI will help design experts by making assumptions in forecasts and capacity planning. AI will simulate various usage scenarios and support design experts by providing suggestions on sizing conductors, transformers, and substations to ensure a robust and reliable supply chain, even when newly introduced needs change or fluctuate. These abilities are crucial as more operations shift to remote settings, where oversizing and undersizing can lead to high costs and underlying reliability issues. These design suggestions will also prevent oversizing and undersizing, while dynamically guiding design experts to make informed decisions that optimize infrastructure performance (Alhamrouni, 2024, p. 32).

AI could provide resilience and reliability analysis for design experts. There are always numerous environmental and operational contingencies to mitigate the failure of the plan due to issues such as wildfire risk, flooding, storms, and equipment failures. Predictive analytics can prioritize maintenance scheduling and provide real-time solutions and operational arrangements, minimizing operational disruption. AI can use dynamic data from the Internet of Things (IoT) to identify the early warning signs that could lead to potential failure or system inefficiencies (Koekkoek, 2025, Para. 3).

By utilizing machine learning algorithms to generate site data, optimize routing, and provide various design scenarios and suggestions, the projects were able to reduce manual inputs and expedite the decision-making process for design experts. The use of AI in CAD has demonstrated significant reductions in design time, particularly in planning and modeling, for transmission projects by eliminating bottlenecks that can emerge in the manual design process. This process also helps designers transition to online platforms more rapidly, which is crucial for meeting increased design demand and enhancing efficiency (Buga, 2025, p. 38).

**Advantages of Artificial Intelligence-Driven CAD Systems**

Artificial intelligence-driven CAD will continue to evolve and develop as different needs emerge, optimizing its processes to meet the evolving requirements of higher efficiency, cost reduction, data extraction, and sustainability. These are critical factors that designers, engineers, and construction firms must assess to ensure effective project planning and execution. The demand for comprehensive analysis is sure to increase over time. AI-driven CAD holds high potential that could significantly expand for innovation, scalability, and functional growth.

As technological advancements continue, the efficiency and design precision of AI algorithms will improve, providing richer, more precise, and more effective results. AI will be better equipped for company users as they use it more and the database grows. AI will be able to provide even more intellectually optimized pipe routing, simulate more accurate hydraulic behavior, and make instant adjustments to terrain or climate changes, thereby reducing planning and modeling timelines. For power transportation, AI will be able to provide even more detailed, complex, and sophisticated route optimization and predictive modeling, allowing faster deployment of transmission and distribution networks while ensuring maximum safety and compliance.

Diagram 3, Different advantages enhancing each other in a cycle.

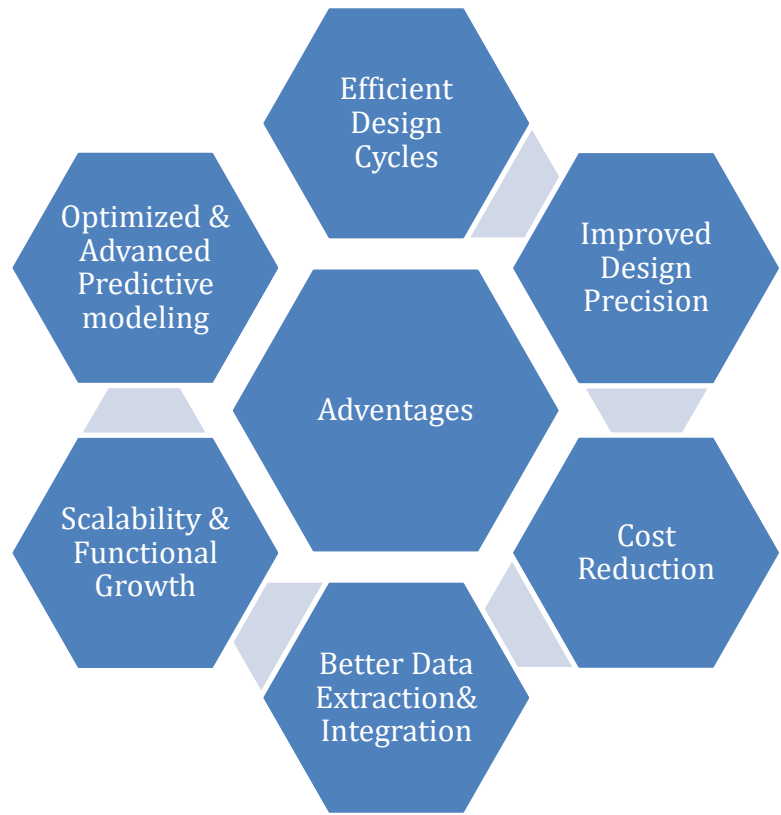

With these upgraded models, the design process will be able to predict complete maintenance needs, resulting in cost reduction. AI will continuously analyze real-time data and learn to provide more advanced solutions as the data grows over time. Real-time data from sensors and historical failure patterns will enable early detection of risks across all aspects. For example, the more pipeline leaks or transformer faults there are, the more the AI will analyze not only the current situation but also data from before and after incidents. This will result in reduced unscheduled downtime and an extended asset lifespan. In the water sector alone, AI reduces operational expenditures by 20-30% through energy optimization, predictive maintenance, and advanced resource allocation (Barazandeh, 2025, Para. 4). With the advancements of AI, the industry can expect even more efficient suggestions that AI provides. These are already saving on the cost of the current operation, and the evolution of AI will save even more in remote or resource-constrained settings.

In the future, data integration and collaboration will facilitate seamless collaboration between software and hardware. Since the relatively recent adoption of AI, there are still many limitations in data integration. The limitations will be resolved as technology and datasets evolve, and another AI will address data-integration limitations, helping close the gaps between disparate data formats and systems. With these advancements, it will enable greater opportunities for many software applications, such as geographic information systems and building information systems, to transform their data into a compatible format and enhance contextual awareness, risk assessment, and lifecycle management, which are critical sources for sustainable infrastructure (Zhu, 2022, p. 12).

Sustainability and resilience can be enhanced with the aid of AI. Using collective data, artificial intelligence will simulate the impact of climate change across different scenarios and present these possibilities to design experts. This information will enable the design of resilient water and power transportation systems. For example, advanced machine learning powered by long-term collected data will predict high-risk flood zones, strategically locate stormwater controls, and create a resilient and effective plan for extreme weather events (Abdulameer, 2025, P. 15). The prediction provided by analyzing collected data will become more precise as the term of data collection increases. It will be crucial information as aging infrastructure and environmental conditions shift, the associated challenges increase, and become more complex.

**Future trends and innovations**

In the future, the integration of AI with CAD is expected to revolutionize infrastructure design and management, driven by various trends and groundbreaking innovations. There are many innovations still in their early stages, with significant potential to reshape the CAD design industry. The ability to automate and detect adaptive control will increase as the demand for asset and crisis management rises. For example, AI will be able to enhance its real-time analysis capabilities through the integration of AI with the Internet of Things, enabling instant detection of abnormalities and adaptive control (Nwadiokwu, 2024, p. 1877). This ability will become more accurate as more data is accumulated over time, and it is particularly valuable, especially for system operations in remote and underdeveloped locations.

The twin technologies simulate every potential situation by replicating real-life scenarios to generate the best options. The evolution of digital twin technology will create enhanced virtual replicas of physical infrastructure and provide improved AI-driven predictive analytics, which will provide more accurate benchmarking, scenario testing, and infrastructure optimization for both water and energy transportation systems (Homaei, 2024, P. 9). Ultimately, the digital twin and AI will be able to facilitate automated decision-making and mitigate risk by themselves.

Another technology that could improve the integration of AI with CAD is smart grids. The smart grid is one of the breakthrough innovations that has advanced electricity networks by utilizing digital technology and two-way communication. This enhances the efficiency of power networks, making them more reliable and sustainable (Balamurugan, 2025, p. 5). Due to technological breakthroughs beyond traditional one-way flow grids, the smart grid enables real-time monitoring, control, and management of power transmission. By integrating the smart grid into AI-driven CAD, there is potential to unlock new technological breakthroughs and reveal previously unexplored possibilities beyond conventional approaches.

**Conclusion**

The study was conducted to examine how integrating artificial intelligence into computer-aided design systems can improve our design workflow in planning, analysis, and development of water and power transportation infrastructure in remote and undeveloped regions. Developing this system is not only building them but also controlling and maintaining them over time.

The examination explored how an AI-driven CAD system can reduce design time, optimize resources, and predict maintenance, while making the process and product more sustainable and operationally resilient. All these aspects are potentially automatable, and automation could be crucial in remote and underdeveloped locations, especially in the design of water and power transportation infrastructure. Furthermore, these advantages can lead to reduced project delivery time and costs, while improving reliability and collaboration among experts and software. It is also essential to address ongoing advancements and evolutions in technologies such as digital twins, smart grids, and real-time data integration, which could pair with AI-driven CAD to revolutionize the design process, thereby increasing the capabilities of these systems by tapping untapped potential. With the AI-driven CAD system, designers can design more efficiently, with greater accuracy, and make better decisions.

In summary, this study set out to examine how integrating AI into CAD systems can address the unique challenges of designing water and power transportation infrastructure in remote and underdeveloped areas, directly reflecting the purpose and research question stated at the outset. The findings demonstrate that AI-driven CAD enables faster project delivery, improved resource optimization, and predictive maintenance, while increasing sustainability and operational resilience. These outcomes align with the paper's initial aim of exploring both the advantages and challenges of AI-CAD integration. As demand for robust infrastructure grows, AI-powered CAD will be essential for meeting the needs of efficiency, cost-effectiveness, and adaptability, as well as for supporting innovations such as GIS and IoT integration. Ultimately, by tying back to the introduction and purpose statement, this research confirms that AI-integrated CAD systems provide effective, forward-looking solutions to present and future infrastructure challenges, enabling the development of resilient, sustainable, and technologically advanced systems.